\documentclass[prd,12pt,aps,preprintnumbers,tightenlines,floatfix,nofootinbib,amssymb]{revtex4}

\newcommand{\beqn}{\begin{eqnarray}}
\newcommand{\eeqn}{\end{eqnarray}}
\newcommand{\eq}[1]{(\ref{#1})}
\newcommand{\Tr}{{\mathrm{Tr}}\,}
\newcommand{\cD}{{\cal D}}
\newcommand{\cF}{{\cal F}}
\newcommand{\cE}{{\cal E}}
\newcommand{\dd}{{\mathrm d}}
\newcommand{\vn}{{\mathbf{n}}}
\newcommand{\vx}{{\mathbf{x}}}

\newcommand{\vnabla}{\mathbf{\nabla}}
\newcommand{\Z}{{Z \!\!\! Z}}
\def\bbbone{{\mathchoice {\rm 1\mskip-4mu l} {\rm 1\mskip-4mu l} {\rm 1\mskip-4.5mu l} {\rm 1\mskip-5mu l}}}
\newcommand{\ITEP}{\affiliation{Institute of Theoretical and
Experimental Physics, B.Cheremushkinskaya 25, Moscow, 117259, Russia}}

\begin{document}

\title{Yang-Mills theory in Landau gauge as a liquid crystal}

\author{M.N.~Chernodub}\ITEP

\preprint{ITEP-LAT/2005-09}

\begin{abstract}
Using a spin--charge separation of the gluon field in the Landau gauge
we show that the $SU(2)$ Yang-Mills theory in the low-temperature phase
can be considered as a nematic liquid crystal. The ground state of the
nematic crystal is characterized by the $A^2$ condensate of the gluon field.
The liquid crystal possesses various topological defects (instantons, monopoles and vortices)
which are suggested to play a role in non-perturbative features of the theory.
\end{abstract}


\date{June 12, 2005}

\maketitle

Separation of degrees of freedom is a useful analytical tool which is widely used in various physical applications.
For example, the spin-charge decomposition (often referred to as the slave-boson formalism~\cite{ref:slave-boson}) of the strongly
correlated electrons is a popular technique invoked to describe a low-temperature physics of the high-$T_c$ cuprate
superconductors~\cite{ref:highTc}. According to the slave-boson formalism the electron creation operator $c^\dagger_{i\sigma}$
is represented as the product of two operators,
\beqn
c^\dagger_{i\sigma} = f^\dagger_{i\sigma} b_i\,,
\label{eq:spin:charge:electron}
\eeqn
where $i$ is the lattice site and $\sigma=\uparrow,\downarrow$ is the spin index.
The operator $f^\dagger_{i\sigma}$ creates a chargeless spin-1/2 fermion state ("spinon") while
the operator $b_i$ annihilates a charged spin-0 boson state ("holon"). Physically, the electron
is represented as a composite of the spinon particle (which carries information about the spin of the
electron) and the holon particle (which knows about the electron charge). The decomposition
conserves the total number of the degrees of freedom because of the constraint
$f^\dagger_{i\uparrow} f_{i\uparrow} + f^\dagger_{i\downarrow} f_{i\downarrow} + b^\dagger_i b_i =1$.
In Eq.~\eq{eq:spin:charge:electron} the states with double occupancy are disregarded for simplicity.

The local nature of the spin-charge decomposition~\eq{eq:spin:charge:electron} leads to an emergence of an
internal {\it compact} $U(1)$ gauge symmetry realized in the form of the gauge transformations
\beqn
f_{i\sigma} \to e^{i \varphi_i} \, f_{i\sigma}\,,\quad b_i \to e^{i \varphi_i} \, b_i\,.
\label{eq:gauge:U1}
\eeqn
Certain properties of the high-$T_c$ superconductors can be described~\cite{ref:highTc,ref:high_Tc-U1}
by $U(1)$ gauge models which are utilizing the mentioned internal gauge symmetry. These gauge models are
treatable within the mean field approach which predicts a rich phase structure. In particular, the
$d$-wave high--$T_c$ superconductor is suggested~\cite{ref:high_Tc-U1} to be realized as a phase in which
the spinon pairing,
$\Delta_{ij} \equiv \langle f^\dagger_{i\uparrow} f_{j\downarrow} - f^\dagger_{i\downarrow} f_{j\uparrow} \rangle \neq 0$,
is accompanied with a spontaneous breaking of the internal $U(1)$ symmetry by the holon condensate:
\beqn
b \equiv \langle b_i \rangle \neq 0\,.
\label{eq:b}
\eeqn
The presence and the subsequent spontaneous breaking of the internal gauge symmetry may have important
physical consequences if even this symmetry is not realized in the original formulation of the theory.

Quantum Chromodynamics is another example of a strongly interacting system in which the breaking of the internal
symmetry may play an essential role. Long time ago it was suggested~\cite{ref:dual:superconductor} that the confinement of
quarks into hadrons may happen due to a condensation of special gluonic configurations called Abelian monopoles. In this approach
-- referred to as the dual superconductivity scenario -- a condensate of the monopoles breaks spontaneously an internal
(or, "dual") $U(1)$ gauge symmetry. According to the dual superconductor idea, the breaking of the dual symmetry gives rise
naturally to the dual Meissner effect, which insures a formation of a QCD string, which in turn leads to the confinement
the quarks into hadronic bound states.

The problem of an explicit realization of the dual superconductivity
in QCD in terms of the original (gluon) fields is not solved yet. Moreover, the dual superconductivity is shown numerically
to be realized~\cite{ref:dual:superconductivity:numerical} only in a special Maximal Abelian gauge which explicitly selects
predefined direction(s) in the color gauge group. In this gauge the gluons from the diagonal (Cartan) subgroup are likely to be
responsible for the infrared phenomena such as the quark confinement~\cite{ref:dual:superconductivity:numerical}. The off-diagonal
gluons were shown to be short--ranged and are largely inessential for the infrared physics~\cite{ref:off:diagonal}.

Recently, it was proposed~\cite{ref:splitting:niemi} to split the
gluons in a manner of the spin-charge separation used
in the high-$T_c$ superconductivity models. The splitting is based
on the field decomposition~\cite{ref:splitting:faddeev} which is
applied to the off-diagonal gluons while leaving the diagonal
gluons intact. In the SU(2) Yang-Mills (YM) theory the splitting
of the off-diagonal gluons~\cite{ref:splitting:niemi,ref:splitting:faddeev},
\beqn
A^1_\mu + i A^2_\mu = \psi_1 \, {\mathbf e}_\mu + \psi^*_2 \, {\mathbf e}^*_\mu,
\hspace{5mm}
{\mathbf e}_\mu {\mathbf e}_\mu = 0,
\hspace{3mm}
{\mathbf e}_\mu {\mathbf e}^*_\mu = 1,
\label{eq:separation:niemi}
\eeqn
leads to appearance of two electrically charged (with respect to the Cartan subgroup
of the color gauge group) Abelian scalar fields
$\psi_{1,2}$ and the electrically neutral field ${\mathbf e}_\mu$ which is a complex vector.
There are also other popular gluon field decompositions~\cite{ref:splitting:cho}, some
of which were suggested to be related to the monopole condensation.

In this paper we propose a novel generalization of the spin-charge decomposition
of the high--$T_c$ superconductors~\eq{eq:spin:charge:electron} to the $SU(2)$ Yang--Mills (YM) theory.
This decomposition splits the $SU(2)$ gluon field into spin and color degrees of freedom treating
all color components equally:
\beqn
A^a_\mu(x) = \Phi^{ai}(x) \, e^i_\mu(x)\,.
\label{eq:separation}
\eeqn
Here $\Phi^{ai}(x)$ is the $3\times 3$ matrix, and $a^i_\mu(x)$ are the three vectors forming an (incomplete)
orthonormal basis in the four dimensional space-time, $e^i_\mu(x) e^j_\mu(x) = \delta^{ij}$.
The elements of $\Phi^{ai}(x)$ and $a^i_\mu(x)$ are real functions
labeled by the color ($a=1,2,3$), internal ($i=1,2,3$) and Euclidean vector ($\mu=1,\dots,4$) indices.
Obviously, Eq.~\eq{eq:separation:niemi} is a color-symmetric generalization of Eq.~\eq{eq:separation:niemi}.
In order to avoid cluttering of notations with lower and upper indices we prefer to work in
the Euclidean space--time.

The splitting~\eq{eq:separation} of the gluon fields can obviously
be written in any SU(2) gauge. However, under the local $SU(2)$
color transformations, $A_\mu(x) \to A^{\Omega}_\mu(x) =
\Omega\bigl(A_\mu + i g\, \partial_\mu\bigr) \Omega^\dagger$,
the fields $\Phi^{ai}$ and $e^i_\mu$ mix with each other in a
complicated way. Here $A_\mu \equiv A^a_\mu \sigma^a/2$ is the
gauge field, $\sigma^a$ are the Pauli matrices, and $g$ is the gauge coupling.

In order to make the splitting~\eq{eq:separation} well--defined we fix the
Landau gauge~\eq{eq:F} which minimizes the gauge fixing functional,
\beqn
\min_\Omega F[A^{\Omega}]\,,\qquad
F[A] = \int \dd^4 x \, {\left[A^a_\mu(x)\right]}^2\,,
\label{eq:F}
\eeqn
over the gauge transformations. This gauge condition fixes the $SU(2)$ color gauge freedom up to
the $SU(2)$ {\it global} freedom (which is usually disregarded):
$A^a_\mu(x) \to \Omega^{ab}_{\mathrm{gl}} \, A^b_\mu(x)$, where $\Omega^{ab}_{\mathrm{gl}} =
\Tr \bigl(\sigma^a \Omega \sigma^b \Omega^\dagger \bigr)/2$
is the coordinate-independent matrix belonging to the adjoint representation of the color $SU(2)$ group.

The transformation rules for the components of the gauge field~\eq{eq:separation} are:
\beqn
\Phi(x) \to \Omega_{\mathrm{gl}} \, \Phi(x) \, \Lambda^T(x)\,, \quad
e_\mu(x) \to \Lambda(x) \, e_\nu(x) \, \xi_{\mu\nu}\,,
\label{eq:transformations}
\eeqn
or, explicitly, $\Phi^{ai}(x) \to \Omega^{ab}_{\mathrm{gl}} \, \Phi^{bj}(x) \, \Lambda^{ij}(x)$ and
$e^i_\mu(x) \to \Lambda^{ij}(x) \,  e^j_\nu(x) \, \xi_{\mu\nu}$.
Here $\Omega_{\mathrm{gl}}$ is the matrix of the SU(2) color global transformations,
$\xi_{\mu\nu}$ is the SO(4) element of the rotations in the Euclidean space-time and $\Lambda$ is the matrix of the internal SO(3)
transformations ($\Lambda^T\Lambda = \bbbone$):
\beqn
\Omega_{\mathrm{gl}} \in SO(3)^{\mathrm{color}}_{\mathrm{global}}\,, \qquad
\xi \in SO(4)^{\mathrm{spin}}_{\mathrm{global}}\,, \qquad
\Lambda(x) \in SO(3)^{\mathrm{internal}}_{\mathrm{local}}\,.
\label{eq:transformations:group}
\eeqn
Equation~\eq{eq:separation} can be interpreted as a spin-color separation of the gluon field
since the color gauge transformations $\Omega_{\mathrm{gl}}$ are acting only on the matrix field $\Phi$ while
spin transformations $\xi$ affect only the field $e_\mu$. Note that the color and the space-time rotations are
the symmetries of the original $SU(2)$ gauge theory while the internal symmetry group originates from the form
of the splitting~\eq{eq:separation} in analogy to the compact $U(1)$ internal symmetry~\eq{eq:gauge:U1} of the
high-$T_c$ superconductors. For the sake of brevity we call below the group of the internal gauge
transformations as $SO(3)_{\mathrm{int}}$.

The proposed splitting~\eq{eq:separation} is self-consistent from the point of view of counting
of the degrees of freedom (d.o.f.). The original field $A^a_\mu$
is described by $3 \times 4 = 12$ real functions\footnote{While counting the degrees of freedom we
do not take into account the pure color gauge
degrees of freedom ($-3$ d.o.f.) and do not impose the Gauss
constraint ($-3$ d.o.f.) to select the physical states because these restrictions equally
affect both sides of Eq.~\eq{eq:separation}.}. The
field $A^a_\mu$ is now rewritten~\eq{eq:separation} as the product
of the matrix $\Phi^{ai}$ ($3 \times 3 = 9$ d.o.f.) and the three
vector fields $e^i_\mu$ ($3 \times 4 = 12$ d.o.f.)
subjected to the orthonormality constraints ($-6$ d.o.f.). The
group $SO(3)_{\mathrm{int}}$ of the internal gauge transformations
has 3 generators ($-3$ d.o.f.). Thus, the number of the d.o.f. in
the field $A^a_\mu$ (which is 12) is the same as the total number
of d.o.f. in the product of the fields $\Phi^{ai}$ and $e^i_\mu$:
(which is $9 + 12 - 6 - 3 = 12$).

It is instructive to rewrite the SU(2) gauge model in an explicitly $SO(3)_{\mathrm{int}}$ invariant form.
To this end one may introduce two composite gauge fields:
\beqn
\Gamma^{ij}_\mu = \frac{1}{2} \bigl(e^i_\nu \partial_\mu e^j_\nu - e^j_\nu \partial_\mu e^i_\nu \bigr) \,, \quad
\vartheta^{ij}_\mu = \frac{1}{2} \Bigl((\Phi^{-1})^{ia} \partial_\mu \Phi^{aj}  -
                     (\Phi^{-1})^{ja} \partial_\mu \Phi^{ai} \Bigr) \,,
\label{eq:def:gauge}
\eeqn
and two composite matter fields,
\beqn
\chi^{ij} = \Phi^{ai} \Phi^{aj}\,,\qquad
z^{ij}_\mu = \frac{1}{2} \Bigl((\Phi^{-1})^{ia} \partial_\mu \Phi^{aj}  +
                     (\Phi^{-1})^{ja} \partial_\mu \Phi^{ai} \Bigr)\,,
\label{eq:def:matter}
\eeqn
which transform under the internal gauge transformations as
\beqn
\Gamma_\mu \to \Lambda (\Gamma_\mu  + \partial_\mu) \Lambda^T\,,
\quad
\vartheta_\mu \to \Lambda (\vartheta_\mu  + \partial_\mu) \Lambda^T\,,
\quad
\chi \to \Lambda \chi \Lambda^T\,,
\quad
z_\mu \to \Lambda z_\mu \Lambda^T\,.
\label{eq:int:gauge:transf}
\eeqn
The $SO(3)_{\mathrm{int}}$ gauge fields $\Gamma_\mu$ and $\vartheta_\mu$ are asymmetric with respect to
permutations of the internal indices while the scalar matter field $\chi$ and the vector matter field $z_\mu$ are
symmetric under these permutations. The matter fields transform in the adjoint representation of
the $SO(3)_{\mathrm{int}}$ gauge group. Note that it is impossible to construct composite matter fields from
the "spin" field $e_\mu$ in a manner of Eq.~\eq{eq:def:matter} due to the orthonormality constraints imposed on $e_\mu$.

The Landau gauge functional~\eq{eq:F} can be expressed in terms of the matter field $\chi$
\beqn
F[A] \equiv \cF[\chi] = \int \dd^4 x \,  \Tr \chi\,.
\label{eq:F:chi}
\eeqn
Note that this functional still invariant under all global and
local transformations~\eq{eq:transformations:group}.

Technically, the existence of the two gauge fields~\eq{eq:def:gauge} and one adjoint vector field~\eq{eq:def:matter} allows
us to define an arbitrary number of covariant derivatives,
$D^{ij}_\mu(\gamma) = \partial_\mu \,\delta^{ij} + \gamma^{ij}_\mu$
where the vector field $\gamma_\mu$ stands for any linear combination of the $\Gamma_\mu$, $\vartheta_\mu$ and $z_\mu$
fields which transforms as a $SO(3)_{\mathrm{int}}$ gauge field. Then, the
derivative of the gauge field $A^a_\mu$ can be represented in an explicitly $SO(3)_{\mathrm{int}}$
invariant form, $\partial_\mu A^a_\nu = (\hat \Phi^a, D_\mu \hat e_\nu) + (D_\mu \hat \Phi^a, \hat e_\nu)$.
The local (differential) condition of the Landau gauge, $\partial_\mu A^a_\mu = 0$, can be rewritten as a
constraint
\beqn
(\hat \Phi^a, D_\mu \hat e_\mu) + (D_\mu \hat \Phi^a, \hat e_\mu) = 0\,.
\label{eq:Landau:gauge:local}
\eeqn
Here vectors $\hat \Phi^a \equiv (\Phi^{a1},\Phi^{a2},\Phi^{a3})^T$ are the columns of the matrix $\Phi^{ai}$,
$\hat e_\mu = {(e^1_\mu,e^2_\mu,e^3_\mu)}^T$, and $(a,b) = a^i b^i$ is the scalar product in the
internal $SO(3)_{\mathrm{int}}$ space. Below we make the choice $\gamma^{ij}_\mu = \Gamma^{ij}_\mu$ for convenience.

It is also convenient to introduce the vector $e^4_{\mu} = \varepsilon_{\mu\nu\alpha\beta}
e^1_\nu e^2_\alpha e^3_\beta$. The four vectors $e^{\bar i}_\mu$, $\bar i=1,\dots,4$ form a complete
orthonormal basis in the $4D$ space-time, $e^{\bar i}_\mu e^{\bar j}_\mu = \delta^{\bar i\bar j}$. The internal
$SO(3)_{\mathrm{int}}$ transformations act in the subspace spanned onto vectors $e^k_\mu$ with $k=1,2,3$ while
leaving the vector $e^4_{\mu}$ intact.

The YM Lagrangian be divided into the three parts
\beqn
L_{SU(2)}[A] \equiv \frac{1}{4} {\left[ G^a_{\mu\nu}(A)\right]}^2
= L_0[\Phi,\chi,\Gamma] + L_1[\chi,\Gamma,\vartheta] + L_2[\chi] + L_{\mathrm{gf}}\,,
\label{eq:YM:standard}
\eeqn
where $G^a_{\mu\nu}(A) = \partial_{[\mu,} A^a_{\nu]} + g\, \varepsilon^{abc} A^b_\mu A^c_\nu$ is
the $SU(2)$ field strength tensor and the term $L_n$ is proportional to the $n^{\mathrm{th}}$ power of the $SU(2)$
coupling constant $g$. For a moment we disregard the term $L_{\mathrm{gf}}$ coming from the
Landau gauge fixing. Using an appropriate multiplication by the vectors $e^{\bar k}_\mu$ to convert the Euclidean
indices into the internal $SO(3)_{\mathrm{int}}$ basis we rewrite the YM Lagrangian~\eq{eq:YM:standard} as follows:
\beqn
L_0[\Phi,\chi,\Gamma]
& = & \frac{1}{2} \Bigl(\cD_{\bar k}(\Gamma) {\hat \Phi}^a\Bigr)^2
+ \frac{1}{2} \Bigl(\Sigma(\Gamma), \chi \Sigma(\Gamma)\Bigr),
\label{eq:L0}\\
L_1[\chi,\Gamma,\vartheta]
& = & 2 g \, \sqrt{{\mathrm{det}}\, \chi} \cdot (\Gamma^{ij}_k - \vartheta^{ij}_k) \, \varepsilon_{ijk}\,,
\label{eq:L1}\\
L_2[\chi]
& = & \frac{g^2}{4} \Bigl[{\left(\Tr \chi\right)}^2 - \Tr {\chi}^2\Bigr]\,.
\label{eq:L2}
\eeqn
where $\cD_{\bar k}(\Gamma) \equiv e^{\bar k}_\mu D_\mu(\Gamma)$ is the covariant derivative acting on
the internal $SO(3)_{\mathrm{int}}$ indices. Note that the spin field $\hat e_\mu$ enters the
Lagrangian~\eq{eq:YM:standard} only in the form of the connection~$\Gamma^{ij}_{\bar k}$.

In order to simplify the $L_0$ part of the YM Lagrangian~\eq{eq:L0} we used the differential Landau gauge condition
and neglected a full-derivative surface term. The first term in $L_0$ is the kinetic term for the "color"
component of the gluon field $\hat \Phi^a$ in the background of the $SO(3)_{\mathrm{int}}$ gauge field $\Gamma$.
The second term in $L_0$ can be interpreted as a "dielectric" energy density
associated with the (space-dependent) "dielectric susceptibility" $\chi$ and the (dynamical) $SO(3)_{\mathrm{int}}$
"electric field" $\Sigma^i(x) = \Lambda^{ij}_\cE(x) \cE_j(x)$. Here the $SO(3)_{\mathrm{int}}$ gauge
transformation $\Lambda^{ij}_\cE$ diagonalizes the matrix
$\Gamma^{4i}_{\bar k} \Gamma^{4j}_{\bar k} =
[\Lambda_\cE \, {\mathrm{diag}}(\cE^2_1,\cE^2_2,\cE^2_3) \, \Lambda^T_\cE]^{ij} $ with $\cE_i(x) \geqslant 0$.

The second part~\eq{eq:L1} of the Lagrangian represents the interaction between the gauge fields
$\Gamma$ and $\vartheta$ with the effective coupling $g \, {\mathrm{det}}^{1/2} \chi \equiv g \, \mathrm{det} \Phi$.
The third part~\eq{eq:L2} is a local potential $V(\chi)$ on the "dielectric susceptibility" field $\chi$.

The analogy of the spin-color separation of the gluon in YM
theory~\eq{eq:separation} with the spin-charge separation of the
electron in the high-$T_c$ superconductor models~\cite{ref:highTc}
manifests itself also in the absence of the kinetic terms for the
composite gauge fields $\Gamma_\mu$ and $\vartheta_\mu$. This fact
is natural since the local construction of each of the composite
gauge fields~\eq{eq:def:gauge} involves already a single
derivative while canonical local Lagrangians ({\it i.e.}, the YM
Lagrangian) contain terms with at most two derivatives. The only explicitly
propagating field in formulation~\eq{eq:YM:standard}~is~$\hat \Phi^a$.

Besides the remarkable analogy of the spin--color separation in the YM theory with the spin-charge separation in the
high--$T_c$ superconductivity, the YM theory has another interesting analogue in the condensed matter physics. Namely, the
YM Lagrangian~(\ref{eq:YM:standard}-\ref{eq:L2}) can be interpreted as the free energy density of a nematic liquid
crystal.

The ordinary nematic crystals~\cite{ref:nematics:review} consist of rod-like molecules which tend to align parallel to
a direction $\vn(\vx, t)$. The molecule is invariant under reflections with respect to a plane perpendicular to
the molecule axis. The unit vector $\vn$ -- called the Frank director -- is  chosen spontaneously in the absence of
external electric or magnetic fields. The molecules in liquid crystals do not have a positional order contrary to solid
crystals characterized by lattice-like structures. The energetically favored ground state of the nematic crystal
is realized at low temperatures and is characterized by a constant director field, $\vn(\vx, t) = \vn_0$.
As temperature increases the system undergoes a transition from the nematic phase to the ordinary (isotropic) phase.

Due to the symmetries of the nematic molecule the symmetry group of the ordinary nematic is $G=SO(3)/\Z_2$.
Therefore, the order parameter in a nematic may be a unit vector but without associated
direction~\cite{ref:nematics:review} ({\it i.e.}, a vector without arrowhead). However, it is more convenient
to define the order parameter to be diadic in $n_i$ similarly to the diamagnetic (or, dielectric)
susceptibility $\tilde \chi_{\alpha\beta}$. The excellent candidate for the order parameter
which discriminates between the nematic and isotropic phases~\cite{ref:nematics:review} is the amount of disorder in
$\tilde \chi_{\alpha\beta}$:
\beqn
{\tilde Q}_{\alpha\beta} = \tilde \chi_{\alpha\beta} - \frac{1}{3} \delta_{\alpha\beta} \, \tilde \chi_{\gamma\gamma}
= \Delta \tilde \chi \sum_{s} \Bigl(n^{(s)}_\alpha n^{(s)}_\beta - \frac{1}{3} \delta_{\alpha\beta}\Bigr) \,,
\label{eq:Q}
\eeqn
where the last equality is written for the molecules with exact axial symmetry. In Eq.~\eq{eq:Q}
the summation is going over all molecules
in a small but macroscopic volume, $\vn^{(s)}$ is the direction of the axis of the $s^{\mathrm{th}}$ molecule, and
$\Delta \tilde \chi = \tilde \chi_\parallel - \tilde \chi_\perp$ is the anisotropy in the diamagnetic (dielectric)
susceptibility along and perpendicular to the molecule axis. The quantity ${\tilde Q}_{\alpha\beta}$ is non-zero in the
nematic phase while it vanishes in the isotropic phase. Below we refer to $\tilde \chi$ as to the dielectric susceptibility.

The dependence of the free energy on the order parameter~\eq{eq:Q} is usually given by an effective Landau--Lifshitz (LL)
potential~\cite{ref:nematics:review},
\beqn
F_{LL}(\tilde Q) = F_0 + \int \dd^3 \vx \, \sum_{n \geqslant 2} \alpha_n \Tr {\tilde Q}^n
\label{eq:F:LL}
\eeqn
where $\alpha_n$ are functions of temperature $T$. The dependence
of the free energy on the isotropic factor $\Tr \tilde \chi$
may be included into the free energy of the normal state, $F_0$.

The deviations of the Frank director $\vn$ from the ground state $\vn_0$ are typically described by the
Oseen--Z\"{o}cher--Frank (OZF) free energy,
\beqn
F_{OZF}[\vn] = \frac{1}{2} \! \int \!\!\dd^3 \vx \Bigl[K_1 (\vnabla \cdot \vn )^2
+ K_2 (\vn \cdot \vnabla \times \vn )^2 + K_3 (\vn \times \vnabla \times \vn)^2\Bigr]\,,
\label{eq:F:OZF}
\eeqn
where the first three terms describe the free energy associated with the splay, twist and bend distortions.
The total free energy of the nematic crystal is $F[{\tilde Q},\vn] = F_{LL}({\tilde Q}) + F_{OZF}[\vn]$.
Note that relation~\eq{eq:Q} makes it possible to rewrite the OZF free energy as a more complicated (compared to \eq{eq:F:OZF})
expression in terms of the order parameter $\tilde Q$.

The YM theory~(\ref{eq:YM:standard}-\ref{eq:L2}) can be associated with
a nematic crystal in which the "molecules" are directed in the internal $SO(3)_{\mathrm{int}}$ space.
There are three species of equivalent molecules in each space-time point
(the number of species equals to the number of the gluons, $N_c=3$).
Consequently, the direction of the local color field in the YM theory,
$(\hat \Phi^a)^i(x)/|\hat \Phi^a(x)|$, is associated with the direction $n^{(a)}_i(x)$
of the $a^{\mathrm{th}}$ molecule species in the point $x$. Then the adjoint matter field
$\chi^{ij} = \sum_a \Phi^{ai} \Phi^{aj}$ can be associated with the dielectric susceptibility,
$\tilde \chi_{\alpha\beta} = \Delta \tilde \chi \sum_{s} n^{(s)}_\alpha n^{(s)}_\beta$.
Note that YM "dielectric susceptibility"  $\chi$ is diadic in the fields $\hat \Phi^a$ similarly to the
dielectric susceptibility $\tilde \chi$ of the nematic.

The proposed association is largely based on the form of the YM term $L_2(\chi)$, Eq.~\eq{eq:L2}, which plays a role of
the LL potential~\eq{eq:F:LL} for the YM "dielectric" field $\chi$. This term can be rewritten via
the isotropic factor $\Tr \chi$ and the traceless symmetric
matrix $Q^{ij}$, constructed from the "susceptibility" $\chi^{ij}$ similarly to the nematic case~\eq{eq:Q}:
$L_2[\chi] = \frac{g^2}{6} (\Tr \chi)^2 - \frac{g^2}{4} \Tr Q^2$.
The negative sign in front of the second term leads to the instability to develop a disorder
in the "dielectric susceptibility" $\chi^{ij}$.

The $L_0$ term, Eq.~\eq{eq:L0}, is a covariant generalization of the kinetic part of the OZF
free energy~\eq{eq:F:OZF} corresponding to the liquid crystal
whose splay, twist and bend distortion constants are equal, $K_1=K_2=K_3=1$. Indeed, in this case the first three terms in
Eq.~\eq{eq:F:OZF} are reduced to $\frac{1}{2} \sum^3_{i,j=1} (\nabla_i n_j)$. Then, we get the $L_0$ term in the YM Lagrangian
by (i) imposing the natural requirement of the $SO(3)_{\mathrm{int}}$ covariance,
$\nabla_\mu \to D_{\bar k}(\Gamma)$, and (ii) taking into account all molecule species, $\vn \to \hat \Phi^{a}(x)$.

As for the $L_1$ term, Eq.~\eq{eq:L1}, it can be interpreted as an energy density associated with a mutual non--alignment
of the directions of the different molecule species $(\hat \Phi^a)^i(x)/|\hat \Phi^a(x)|$.

Let us find the ground state of the nematic associated with the YM theory~(\ref{eq:L0},\ref{eq:L1},\ref{eq:L2}).
In terms of the eigenvalues of the matrix $\chi = {\mathrm{diag}}(\chi_1,\chi_2,\chi_3)$, the ground
state $\chi = \chi^{(0)}$ is defined by the relations:
\beqn
\sum_{\stackrel{i,j=1}{i>j}}^3 \chi^{(0)}_i \chi^{(0)}_j =0 \,,
\quad
\sum_{i=1}^3 \chi^{(0)}_i \geqslant 0\,,
\quad
\prod_{i=1}^3 \chi^{(0)}_i \geqslant 0\,,
\label{eq:ground:state}
\eeqn
where the first relation comes from the condition $\Tr \chi^2 = (\Tr \chi)^2$
corresponding the global minimum of the Ginzburg--Landau potential~\eq{eq:L2}.
The last two relations in Eq.~\eq{eq:ground:state} come from the specific definition of the $\chi$--field~\eq{eq:def:matter}
implying that $\Tr \chi \equiv \sum_{ai} (\Phi^{ai})^2 \geqslant 0$ and ${\mathrm{det}} \chi \equiv ({\mathrm{det}} \Phi)^2 \geqslant 0$,
respectively. Equations~\eq{eq:ground:state} imply that at least two eigenvalues of $\chi$ must be zero. Without loss of
generality we take $\chi^{(0)}_1 = \chi^{(0)}_2 = 0$, and therefore the ground state is $\chi^{(0)} = {\mathrm{diag}}(0,0,\chi_0)$,
where $\chi_0 \geqslant 0$ is not fixed.

The perturbative vacuum (in terms of the original gluon fields
$A^a_\mu$) corresponds to $\chi_0 = 0$, {\it i.e.} to the
isotropic liquid state. What makes the YM field similar to the
nematic liquid is the non--perturbative part of $\chi_0$, which is
fixed by the minimum of the Landau gauge functional~\eq{eq:F:chi}.
This minimum is nothing but the
$A^2$--condensate~\cite{ref:A2:condensate:general}, $\Tr \chi =
\langle A^2_\mu\rangle$, evaluated in the Landau gauge. Thus, the
isotropic liquid state is broken to the nematic crystal state by the
$A^2$ condensate. This spontaneous symmetry breaking of the
isotropic $SO(3)_{\mathrm{int}}$ is similar to the breaking of the
compact gauge group by the holon condensate~\eq{eq:b}.

Technically, a particular non-zero value of the $A^2$-condensate
emerges due to the presence of the gauge--fixing term
$L_{\mathrm{gf}}$ in Eq.~\eq{eq:YM:standard} which also
contributes to the free energy of the nematic liquid and which was
disregarded till now. According to the numerical calculations of
the $A^2$ condensate~\cite{ref:A2:numerical}, $g^2 \chi_0 \approx
(3\,\mbox{GeV})^2$.

The non-perturbative vacuum state, $\chi^{(0)} = {\mathrm{diag}}(0,0,\chi_0)$ with $\chi_0>0$, is still invariant under
the (unbroken) group of rotations about the third axis in the internal space, $H=SO(2)_{\mathrm{int}}$. Due to the fact that
the $SO(3)_{\mathrm{int}}$ gauge field $\Gamma$ is non-propagating, the partial spontaneous breaking of the original internal
symmetry does not lead to a massless vector field.

The interesting question is a possible existence of topological defects
which are generally characterized by non-trivial homotopic groups $\pi_n(G/H)$ of the vacuum manifold $G/H$ of the model.
The vacuum manifold of the YM theory with Lagrangian written in the form~(\ref{eq:YM:standard}-\ref{eq:L2})
is similar to the vacuum manifold of an ordinary nematic~\cite{ref:nematics:topological}
with $G/H = SO(3)/(\Z_2 \times SO(2))$. In particular, the nematic state
contains the $Z_2$ vortices since $\pi_1(G/H) = \Z_2$. This feature may make the physics of the YM nematic state similar
to the center vortex picture of the quark confinement in the YM theory~\cite{ref:center:vortex}.

Moreover, the nematic crystal contains
monopole-like defects characterized by non-negative integers since $\pi_2(G/H) = \Z/\Z_2 \equiv Z_+ = 0, 1,2, \dots$. The monopoles
have the hedgehog--like structure constructed from the arrowless "molecules" (the last fact leads to an identification of the
monopoles with anti-monopoles). The presence of the monopoles may provide a relation between the nematic liquid crystal and the
dual superconductor in the YM theory~\cite{ref:dual:superconductor}. A signature of this relation may already be found in
Ref.~\cite{ref:suzuki} by observing the dual Meissner effect in the Landau gauge.
Finally, the third homotopy group of the vacuum manifold
is also nontrivial, $\pi_3(G/H) = \Z$, which may have a link to the instanton physics.

The disorder, caused by the presence of the described topological defects in the Landau gauge may lead to the non--trivial
consequences for the non--perturbative physics of the YM theory similarly to the effects caused by the center vortex
percolation~\cite{ref:center:vortex} and by the Abelian monopole condensation~\cite{ref:dual:superconductor}.

Finally, we note that lattice
simulations~\cite{ref:A2:condensate:temperature} indicate that the
$A^2$ condensate drops by amount of 92\% at the finite-temperature
phase transition, $T=T_c$. Therefore one may expect that in the
deconfinement phase, $T>T_c$, the $4D$ nematic state may transform
to a $3D$ nematic state characterized by much lower value of the
"nematic dielectric susceptibility" $\chi$. Since the spatial
dynamics of the gluon fields remains non--perturbative in the
deconfinement phase, one may expect that the nematic crystal
splits into two modes: the temporal components of the gluon
fields form an ordinary "isotropic liquid" while the spatial
components are still in a nematic state.

Summarizing, the spin-charge separation idea --
originally invented to describe properties of the high-$T_c$
superconductors -- may also be applied to the YM theory in the
form of the spin-color separation. This approach allows to identify the
ground state of the low-temperature phase YM theory in the Landau gauge
with a nematic liquid crystal. The perturbative isotropic liquid state
is broken down to the nematic liquid crystal state by the $A^2$ condensate.
The nematic crystal contains various topological defects which may play a
role in explaining of non-perturbative features of the YM theory.

\begin{acknowledgments}
The author is supported by grants RFBR 04-02-16079 and MK-4019.2004.2.
The author is grateful to F.V.~Gubarev, A.~Niemi and M.I.Polikarpov for useful discussions.
\end{acknowledgments}

\end{document}